\documentclass[twoside,twocolumn,9pt]{article}
\usepackage{extsizes}
\usepackage[super,sort&compress,comma]{natbib} 
\usepackage[version=4]{mhchem}
\usepackage[left=1.5cm, right=1.5cm, top=1.785cm, bottom=2.0cm]{geometry}
\usepackage{balance}
\usepackage{mathptmx}
\usepackage{sectsty}
\usepackage{graphicx} 
\usepackage{lastpage}
\usepackage[format=plain,justification=justified,singlelinecheck=false,font={stretch=1.125,small,sf},labelfont=bf,labelsep=space]{caption}
\usepackage{siunitx}
\usepackage{float}
\usepackage{fancyhdr}
\usepackage{fnpos}
\usepackage[english]{babel}
\addto{\captionsenglish}{%
  
}
\usepackage{array}
\usepackage{droidsans}
\usepackage{charter}
\usepackage[T1]{fontenc}
\usepackage[usenames,dvipsnames]{xcolor}
\usepackage{setspace}
\usepackage[compact]{titlesec}
\usepackage{hyperref}
\usepackage{stfloats}

\newcommand{\m}[1]{\mathrm{#1}}

\usepackage{epstopdf}

\definecolor{cream}{RGB}{222,217,201}

\begin{document}

\pagestyle{fancy}
\thispagestyle{plain}
\fancypagestyle{plain}{
\renewcommand{\headrulewidth}{0pt}
}

\makeFNbottom
\makeatletter
\renewcommand\LARGE{\@setfontsize\LARGE{15pt}{17}}
\renewcommand\Large{\@setfontsize\Large{12pt}{14}}
\renewcommand\large{\@setfontsize\large{10pt}{12}}
\renewcommand\footnotesize{\@setfontsize\footnotesize{7pt}{10}}
\makeatother

\renewcommand{\thefootnote}{\fnsymbol{footnote}}
\renewcommand\footnoterule{\vspace*{1pt}%
\color{cream}\hrule width 3.5in height 0.4pt \color{black}\vspace*{5pt}} 
\setcounter{secnumdepth}{5}

\makeatletter 
\renewcommand\@biblabel[1]{#1}            
\renewcommand\@makefntext[1]%
{\noindent\makebox[0pt][r]{\@thefnmark\,}#1}
\makeatother 
\renewcommand{\figurename}{\small{Fig.}~}
\sectionfont{\sffamily\Large}
\subsectionfont{\normalsize}
\subsubsectionfont{\bf}
\setstretch{1.125} 
\setlength{\skip\footins}{0.8cm}
\setlength{\footnotesep}{0.25cm}
\setlength{\jot}{10pt}
\titlespacing*{\section}{0pt}{4pt}{4pt}
\titlespacing*{\subsection}{0pt}{15pt}{1pt}

\fancyfoot{}
\fancyfoot[LO,RE]{\vspace{-7.1pt}\includegraphics[height=9pt]{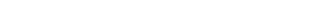}}
\fancyfoot[CO]{\vspace{-7.1pt}\hspace{11.9cm}\includegraphics{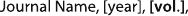}}
\fancyfoot[CE]{\vspace{-7.2pt}\hspace{-13.2cm}\includegraphics{head_foot/RF}}
\fancyfoot[RO]{\footnotesize{\sffamily{1--\pageref{LastPage} ~\textbar  \hspace{2pt}\thepage}}}
\fancyfoot[LE]{\footnotesize{\sffamily{\thepage~\textbar\hspace{4.65cm} 1--\pageref{LastPage}}}}
\fancyhead{}
\renewcommand{\headrulewidth}{0pt} 
\renewcommand{\footrulewidth}{0pt}
\setlength{\arrayrulewidth}{1pt}
\setlength{\columnsep}{6.5mm}
\setlength\bibsep{1pt}

\makeatletter 
\newlength{\figrulesep} 
\setlength{\figrulesep}{0.5\textfloatsep} 

\newcommand{\topfigrule}{\vspace*{-1pt}%
\noindent{\color{cream}\rule[-\figrulesep]{\columnwidth}{1.5pt}} }

\newcommand{\botfigrule}{\vspace*{-2pt}%
\noindent{\color{cream}\rule[\figrulesep]{\columnwidth}{1.5pt}} }

\newcommand{\dblfigrule}{\vspace*{-1pt}%
\noindent{\color{cream}\rule[-\figrulesep]{\textwidth}{1.5pt}} }

\makeatother

\twocolumn[
  \begin{@twocolumnfalse}
{\includegraphics[height=30pt]{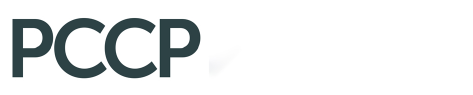}\hfill\raisebox{0pt}[0pt][0pt]{\includegraphics[height=55pt]{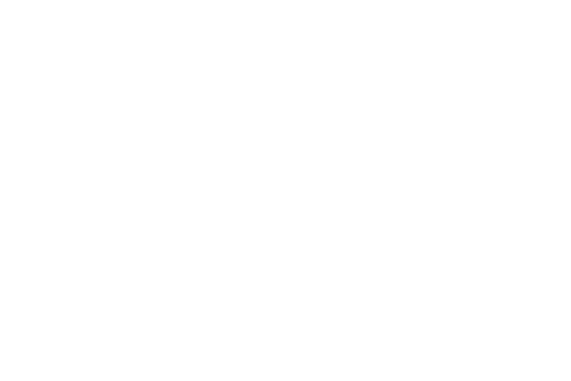}}\\[1ex]
\includegraphics[width=18.5cm]{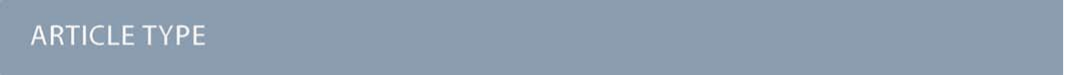}}\par
\vspace{1em}
\sffamily
\begin{tabular}{m{4.5cm} p{13.5cm} }

\includegraphics{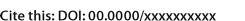} & \noindent\LARGE{\textbf{Detailed Kinetic Model for Combustion of \ce{NH3}/\ce{H2} Blends}}\\
\vspace{0.3cm} & \vspace{0.3cm} \\

 & \noindent\large{Yu-Chi Kao,\textit{$^{a,b}$} Anna C. Doner,\textit{$^{a}$} Timo T. Pekkanen,\textit{$^{a}$} Chuangchuang Cao,{$^{a}$} Sunkyu Shin,\textit{$^{a}$} Alon Grinberg Dana,\textit{$^{c}$} Yi-Pei Li,\textit{$^{b}$} and William H. Green\textit{$^{a}$}}\\

\includegraphics{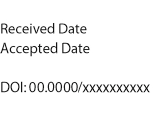} & 
Ammonia is a promising zero-carbon fuel for industrial and transport applications, but its combustion is hindered by flame instabilities, incomplete oxidation, and the formation of nitrogen oxides.  Accurate and detailed kinetic models are critical for designing optimal burners and engines. Despite numerous mechanisms published in recent years, large discrepancies remain between model predictions and experimental data, particularly for \ce{NO_x} species. In this work, we have reviewed the literature to obtain the most up-to-date and reliable thermochemical and kinetic parameters for most reactions present in ammonia combustion, and for reactions for which these parameters are not available, we performed high-level calculations to determine them. The purpose of this was to minimize the number of estimated parameters used in model development. A new, detailed kinetic mechanism was then generated with the Reaction Mechanism Generator (RMG). To ensure physical consistency, geometry optimizations were carried out for all hypothesized “edge” species, and any non-convergent or non-physical structures were excluded. The resulting mechanism was tested against experimental laminar burning velocities, ignition delay time, flow reactor species profiles, and jet-stirred reactor data, and compared with five recent representative mechanisms. Recently developed bath-gas-mixture rules were applied to a number of key reactions in the mechanism, and we found this to result in better agreement with experiment for a number of modeling targets. While the mechanism does not reproduce all experimental results, it demonstrates improved robustness without parameter tuning, thereby reducing the risk of over-fitting and enhancing predictive reliability under conditions relevant to practical applications.

\\

\end{tabular}
 \end{@twocolumnfalse} \vspace{0.6cm}
  ]

\renewcommand*\rmdefault{bch}\normalfont\upshape
\rmfamily
\section*{}
\vspace{-1cm}


\footnotetext{\textit{$^{a}$~Department of Chemical Engineering, Massachusetts Institute of Technology, Cambridge, MA 02139 E-mail: whgreen@mit.edu}}
\footnotetext{\textit{$^{b}$~Department of Chemical Engineering, National Taiwan University, Taipei, Taiwan}}
\footnotetext{\textit{$^{c}$~Department of Chemical Engineering, Technion, Haifa, Israel}}

\footnotetext{\dag~Electronic Supplementary Information (ESI) available: [A PDF file with: details of any supplementary information available should be included here]. See DOI: 10.1039/cXCP00000x/}





\section{Introduction}
Ammonia (\ce{NH3}) has emerged as a promising carbon-free alternative to conventional fuels for industrial and transport applications owing to its intermediate specific energy, ease of handling, and potential for large-scale production from sustainable sources.\cite{kobayashi_science_2019, valera-medina_ammonia_2018} Despite these advantages, its practical use as a fuel poses several challenges. Ammonia is inherently difficult to ignite and sustain in combustion due to its low reactivity, which often results in unstable flames and unacceptable \ce{NH3} emissions.\cite{alnajideen_ammonia_2024} Incomplete oxidation further leads to the formation of pollutants such as nitrogen monoxide and dioxide ($\ce{NO}_x$), raising environmental concerns. To achieve stable and clean operation, ammonia is frequently co-fired with more reactive fuels such as hydrocarbons or \ce{H2}\cite{SZANTHOFFER2023100127}, but this strategy introduces additional design complexity. Accurate and predictive chemical kinetic mechanisms are therefore essential for guiding computational fluid dynamics (CFD) simulations and enabling the design of optimized reactor geometries for minimal environmental impact.

Although numerous \ce{NH3} combustion mechanisms have been developed, recent review articles\cite{DANA2024, MONGEPALACIOS2024101177, SZANTHOFFER2023100127, GIRHE2024113560, huang_uncertainty_2025} have highlighted major inconsistencies between different mechanisms, as well as discrepancies between simulation results and experimental predictions. These limitations stem not only from uncertainties in thermochemical and kinetic parameters, but also from missing key reactions (mechanism-truncation error). In addition, essentially all existing mechanisms neglect the bath-gas-composition dependence of pressure-dependent reactions, or treat it only through the crude classical mixture rule. As demonstrated by Burke and co-workers,\cite{burke_evaluating_2017, lei_bath_2019, singal_implementation_2024} these approximations can have a considerable effect on critical modeling targets like laminar burning velocities (LBVs) and ignition delay times (IDTs).

In this work, we address these shortcomings by 
\begin{itemize}
    \item[1)] updating the Reaction Mechanism Generator (RMG) database with thermochemical and kinetic parameters recommended by Grinberg, Dana and co-workers. \cite{DANA2024, Grinberg2025}
    \item[2)] constructing a mechanism with RMG to reduce mechanism-truncation error. 
    \item[3)] replacing RMG-estimated rate coefficients for reactions with values obtained from transition-state theory (TST) calculations.
    \item[4)] applying improved mixture rules developed by the Burke group for key pressure-dependent reactions.\cite{singal_implementation_2024}
\end{itemize}
The resulting mechanism is very comprehensive and can be used to identify the kinetically important reactions over a wide range of temperatures, pressures, and equivalence ratios relevant to ammonia and hydrogen combustion. Once validated, it can be systematically reduced using model-reduction techniques\cite{echekki_turbulent_2011}, thereby enabling its implementation in computational fluid dynamics (CFD) simulations for practical reactor and burner design.

RMG generates kinetic mechanisms by iteratively selecting species and reactions from a hypothesized ``edge'' and prompting them to the ``core'' mechanism by analyzing computed instantaneous fluxes.\cite{RMG2021, RMG2022} Both edge and core species and reactions require thermochemical and kinetic parameters, for which RMG uses literature values when available and group-additivity and rate-rule estimates when not. In this work, we made a conscious effort to minimize the number of reactions for which RMG needs to rely on such estimates. While this automated approach ensures broad coverage, uncertainties in estimates and the occasional generation of “non-physical” molecules that have valid Lewis structures but are not a minimum on any potential-energy surface (PES) can degrade model accuracy.\cite{Dana2022} To mitigate these issues, we carefully screen the generated species, remove non-physical candidates, and recalculate thermochemical parameters for valid species. This procedure enhances the fidelity of the RMG-generated mechanism.

To further assess performance, we compare in this work our RMG-generated mechanism against state-of-the-art literature mechanisms. Szanthoffer et al.\cite{SZANTHOFFER2023100127} and Girhe et al.\cite{GIRHE2024113560} recently ranked and evaluated \ce{NH3}/\ce{H2} combustion mechanisms according to their agreement with experimental data, identifying both strengths and weaknesses across mechanisms. Here we choose to compare our mechanism with five of the best-performing mechanisms,\cite{STAGNI2023144577, ZHANG2021111653, ZHANG2023127676, MONGEPALACIOS2024101177, ZHU2024113239} as well as experimental results obtained for pure \ce{NH3}, pure \ce{H2}, and \ce{NH3}\ce{H2} mixtures.

\section{Computational and modeling Methods}

\subsection{Rate-Coefficient Calculations}
A detailed description of the rate-coefficient calculations is provided in the ESI\dag, and only a brief summary of the employed methods and programs is given here. Depending on the reaction, geometry optimizations and frequency analyses were performed at the $\omega$B97X-D/def2-QZVP, UHF-CCSD(T)/aug-cc-pVTZ, CASPT2/aug-cc-pVTZ, and $\omega$B97X-D/def2-TZVP levels of theory.\cite{g16,def2basis, WB97XD, dunning} Single-point energies were evaluated at the CASPT2/CBS (complete basis set), CCSD(T)/CBS, and CCSD(T)-F12/cc-pVTZ-F12 levels of theory. Molpro,\cite{molpro1,molpro2,MOLPRO_brief} Gaussian 16,\cite{g16} and ORCA\cite{orca} software packages were utilized to run these calculations. In some cases, we used Arkane\cite{DANA2023} interfaced with Gaussian and Molpro to facilitate the running of quantum-chemistry calculations. CCSDT(Q) corrections were computed with the \textsc{MRCC} program.\cite{mrcc1,mrcc2} Specifically, high-level electronic-structure calculations were performed for the \ce{O(^3P) + HNO}, \ce{^2HO2 + ^3NH}, and \ce{H + ^2H2NO} reaction systems, as detailed in the ESI\dag. Phenomenological rate coefficients were determined with the Automatic Rate Calculator (ARC),\cite{ARC} Arkane,\cite{DANA2023}, and MESMER 7.1.\cite{MESMER} For barrierless reactions, we employed both Gaussrate\cite{gaussrate} and a custom implementation of the Flexible-Transition-State Theory (FTST) method of Robertson et al.;\cite{FTST} this code is provided in the ESI\dag.

\subsection{Thermochemistry Calculations}
For each edge species generated by RMG, geometries were optimized at the $\omega$B97X-D/def2-TZVPD level of theory. Initial guess geometries were obtained by selecting the lowest-energy conformer from a UFF force-field search performed with RDKit.\cite{RDKit} Species with geometries that match the intended graph representation were considered physical. For species that failed this initial step, RDKit was used to generate ten additional conformers, each of which was optimized at the $\omega$B97X-D/def2-TZVPD level of theory. If at least one of these geometries matched the intended graph representation, the species was retained; otherwise, the species was deemed non-physical. In total, 61 species were identified as non-physical. We confirmed that these species do not impact core rate coefficients and forbade these structures during mechanism generation. 

For the remaining species, thermochemical parameters were calculated using ARC. Conformer searches were performed initially at the $\omega$B97X-D/def2-SVP level of theory, with final geometry-optimization, frequency, and hindered-rotor calculations performed at the $\omega$B97X-D/def2-TZVP level of theory. Single-point energies were obtained at the DLPNO-CCSD(T)-F12/cc-pVTZ-F12 level of theory. Atomization energy corrections and Petersson-type bond additivity corrections were applied following the Wu et al. protocol.\cite{WU2024}

\subsection{Mechanism Generation with RMG}
The mechanism was generated under 24 conditions that covered the ranges 700--1900~K, 1.0--100.0~bar, equivalence ratios of $\phi =$ 0.25--2.0, and \ce{NH3}/\ce{H2} blend ratios ranging from 100\%/0\% to 0\%/100\%. Pressure-dependent rate coefficients were estimated using the modified strong collision method with a grain size of 2.1 \si{kJ.mol^{-1}}, over a temperature range of 500–2000~K and a pressure range of 0.1--100~bar. The complete RMG input file, including further details such as selected tolerances, is provided in the ESI\dag.

\subsection{modeling Methods and Analysis}
The generated mechanism for the blended oxidation of  \ce{NH3} and \ce{H2} was tested against a range of experimental data using simulations performed with Cantera 3.1.\cite{cantera} IDTs and \ce{^2OH} time histories from shock tube experiments were compared against results from zero-dimensional (0D) transient simulations. LBVs were determined with one-dimensional (1D) simulations with multi-component transport and thermal diffusion, and then compared with experimental flame-speed data. Finally, the model’s species predictions were tested against concentration-time profiles from jet-stirred reactor (JSR) and flow reactor (FR) experiments; the FR was modeled as a 1D, constant-pressure reactor with the measured centerline temperature profiles as inputs. 

The rate coefficients of complex-forming reactions depend on both pressure and bath-gas composition, but as mentioned in the \emph{Introduction}, the composition dependence is typically neglected by assuming the bath gas is pure \ce{N2} or is approximated with the classical mixture rule. The classical mixture rule is, however, a fairly crude approximation that can introduce errors of up to a factor of two. More worryingly, the work of Burke and co-workers \cite{burke_evaluating_2017, lei_bath_2019, singal_implementation_2024} have shown that the errors introduced by the classical mixture rule tend to be the largest when mole fractions of more efficient colliders such \ce{H2O} and \ce{NH3} are in the 0.1--0.3 range---a common occurrence in ammonia combustion. Fortunately, the mixture rules developed by the Burke group have been implemented in Cantera 3.1, so to improve our treatment of bath-gas-mixture effects, we applied these rules for the following key reactions:
\begin{align}
\ce{^2H + ^2OH &<=> H2O}  \tag{R1} \\
\ce{^2H + ^3O2 &<=> ^2HO2} \tag{R2} \\
\ce{H2O2  &<=> 2 ^2OH} \tag{R3} \\
\ce{NH3  &<=> ^2H + ^2NH2} \tag{R4} \\
\ce{2 ^2NH2  &<=> N2H4}  ~~\text{and} \tag{R5} \\
\ce{HNO  &<=> ^2H + ^2NO} \;.  \tag{R6}
\end{align}
We took the rate-coefficient parameterisations for these reactions from the recent \ce{H2}/\ce{NH3} mechanism from the Burke group.\cite{singal_implementation_2024}

\section{Results and discussion}
\subsection{Rate Coefficients}
\begin{figure}[H]
    \centering
\includegraphics[width=0.9\linewidth]{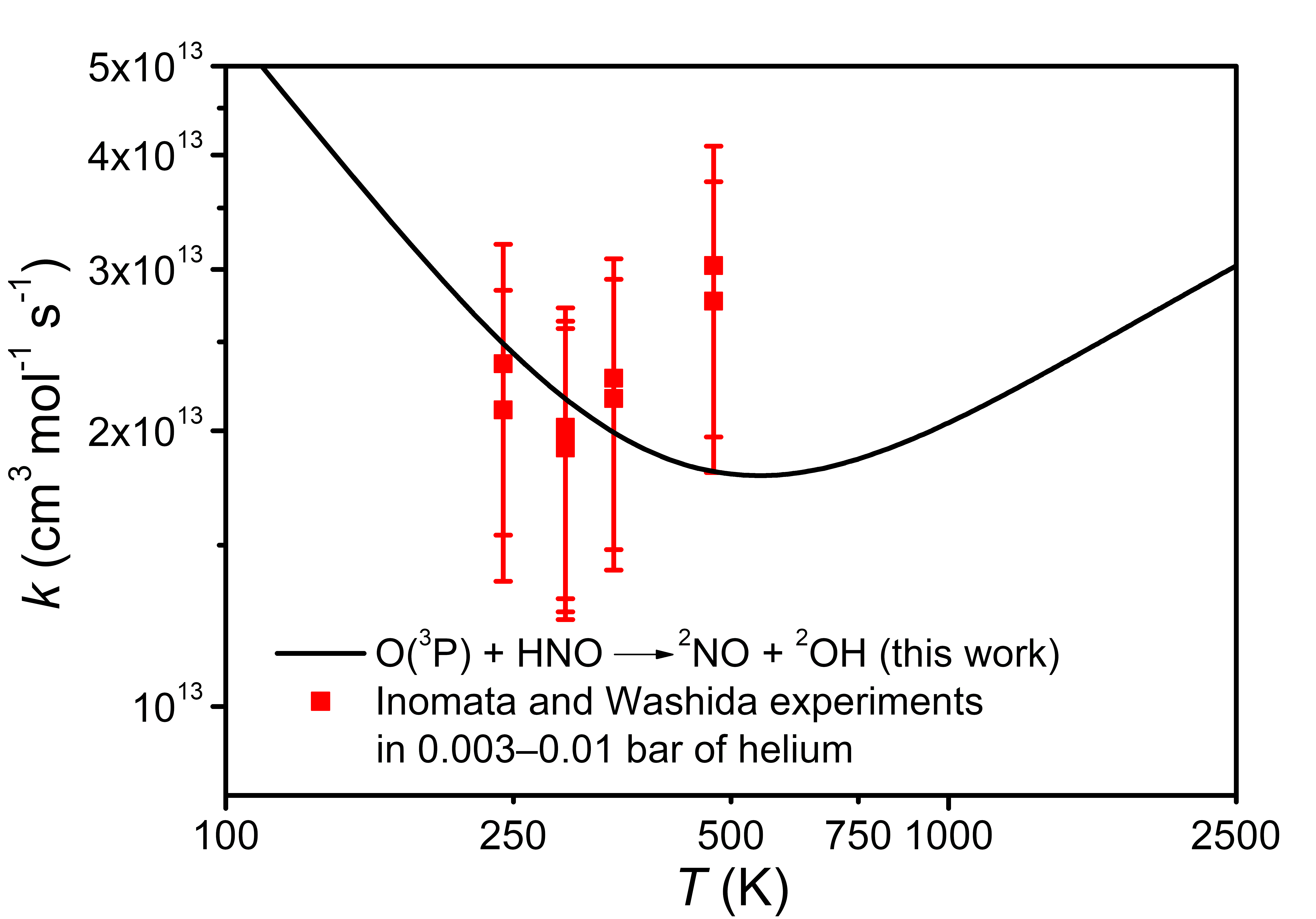}\\
    \caption{Computed rate coefficients for the \ce{O(^{3}P) + HNO -> ^{2}OH + ^{2}NO} reactions shown together with the experimental determination by Inomata and Washida.\cite{INOMATAWASHIDA1999}. The submerged saddle-point for this reaction was lowered by 2.4 \si{kJ.mol^{-1}} to obtain better agreement with experiment.}
        \label{HNO2Fig}
\end{figure}

Table \ref{ratecoefficienttable} lists the reactions added into the RMG-database, as well as their modified Arrhenius parameters. For most of these rate coefficients, there are no experimental measurements, but the 
\begin{align}
\ce{O(^{3}P) + HNO -> ^{2}OH + ^{2}NO}  \tag{R7}
\end{align}
reaction is one exception. We show in Fig. \ref{HNO2Fig} the computed rate coefficient together with the experimental data measured by Inomata and Washida.\cite{INOMATAWASHIDA1999} To obtain better agreement with experiment, we lowered the energy of the rate-determining saddle-point by 2.4 \si{kJ.mol^{-1}}, which is well within computational uncertainties. Master-equation simulations revealed that the recombination reaction to form \ce{^2HNO2} followed by decomposition to \ce{^2H + ^2NO2} is never competitive with R7.

\begin{figure}[H]
    \centering
\includegraphics[width=0.9\linewidth]{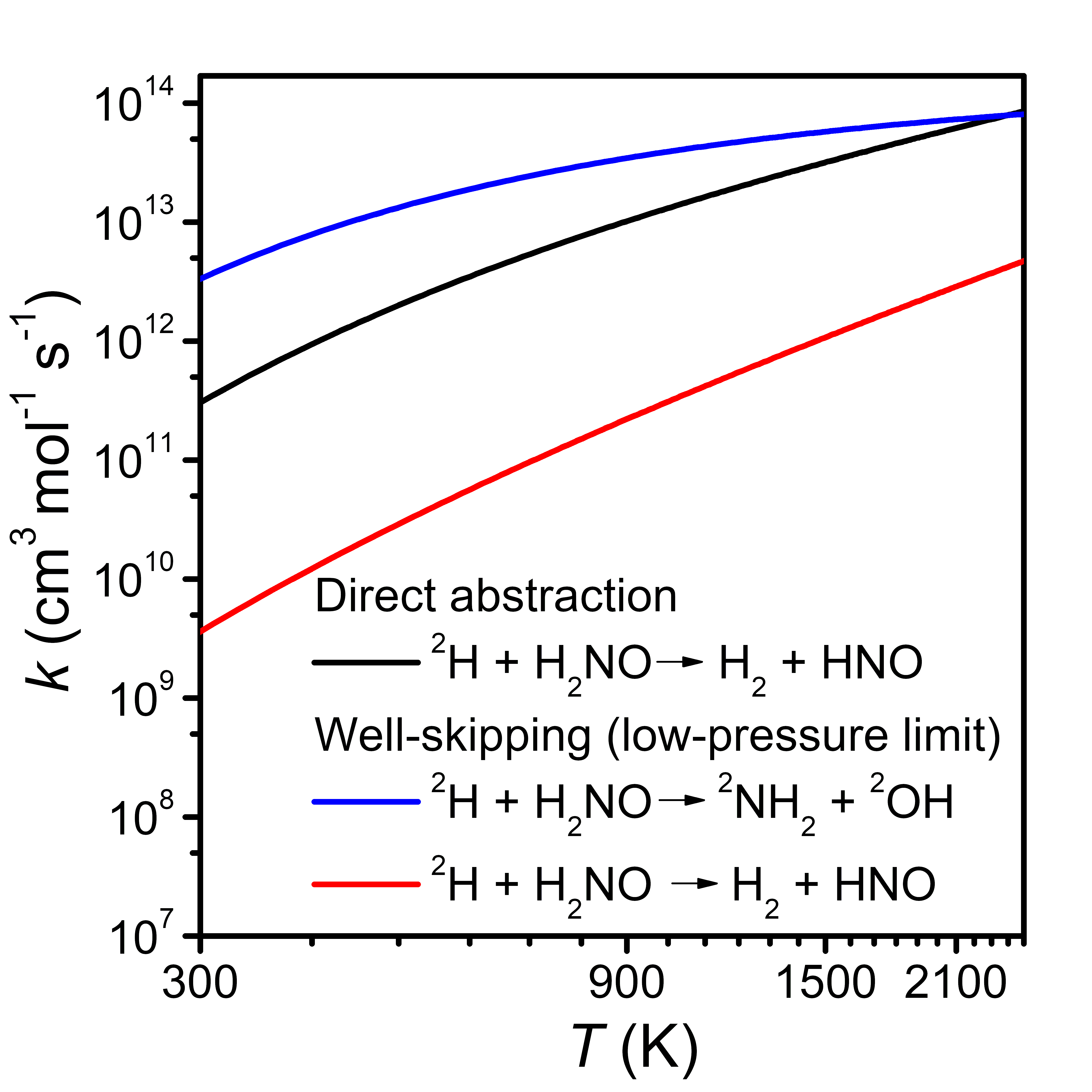}\\
    \caption{Computed rate coefficients for the well-skipping \ce{^2H + ^2H2NO -> (NH2OH/NH3O) -> H2 + HNO} and \ce{^2H + ^2H2NO -> (NH2OH/NH3O) -> ^2NH2 + ^2OH} reactions shown together with the direct-abstraction reaction \ce{^2H + ^2H2NO -> H2 + HNO}. The first two reactions were found to still be at the low-pressure limit at 100~bar}
        \label{H2NOFig}
\end{figure}

\begin{figure}[H]
    \centering
\includegraphics[width=0.9\linewidth]{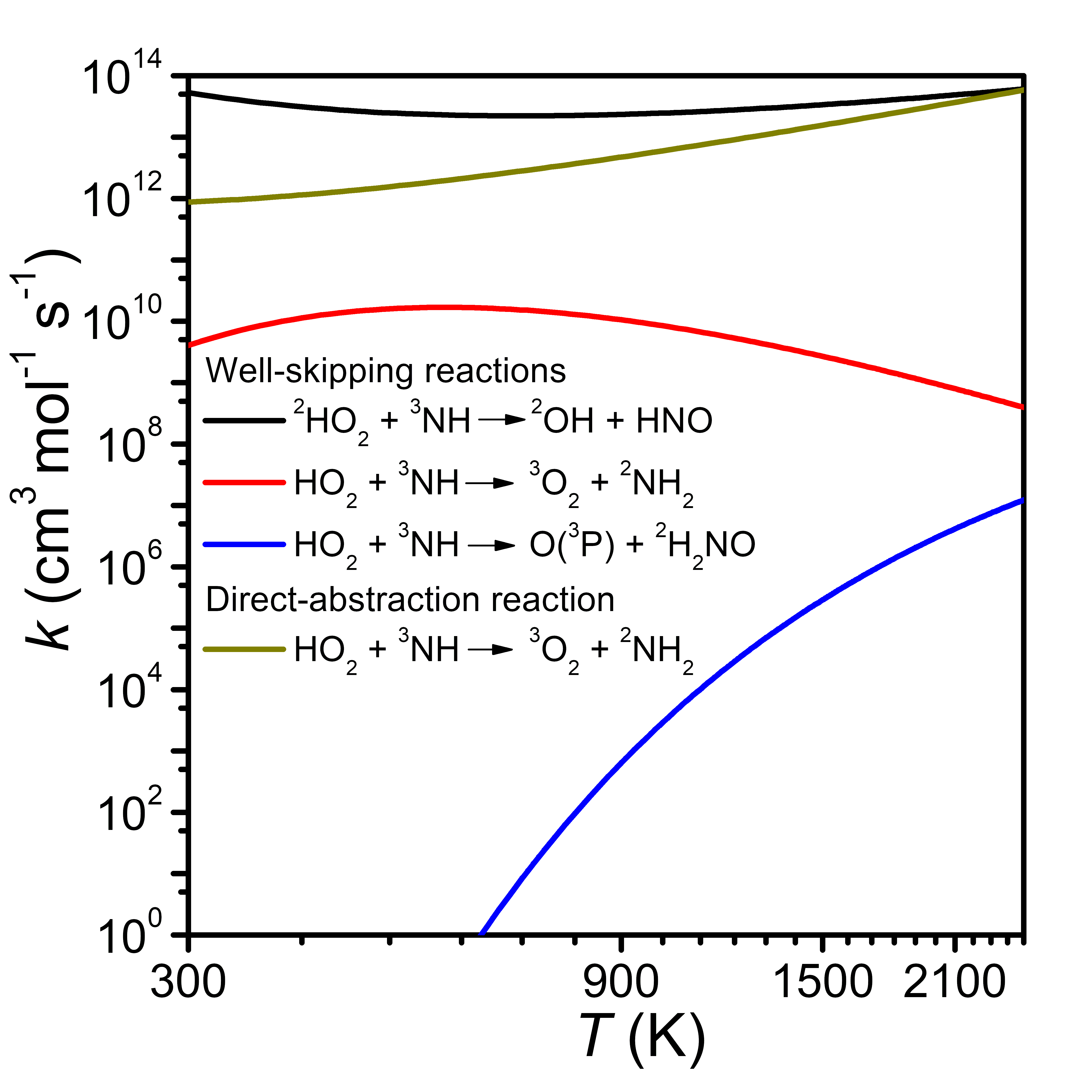}\\
    \caption{Computed rate coefficients for the well-skipping \ce{^2HO2 + ^3NH -> (^2HNOOH) -> ^2OH + HNO}, \ce{^2HO2 + ^3NH -> (^2HNOOH/^2H2NOO) -> ^3O2 + ^2NH2} and \ce{^2HO2 + ^3NH -> (^2HNOOH/^2H2NOO) -> ^3O + ^2H2NO}. The reactions were found to still be at the low-pressure limit at 100~bar. Only the \ce{^2OH + HNO} forming channel is relevant in practice.}
        \label{NHOOHFig}
\end{figure}

\begin{figure*}
    \centering
\includegraphics[width=0.99\linewidth]{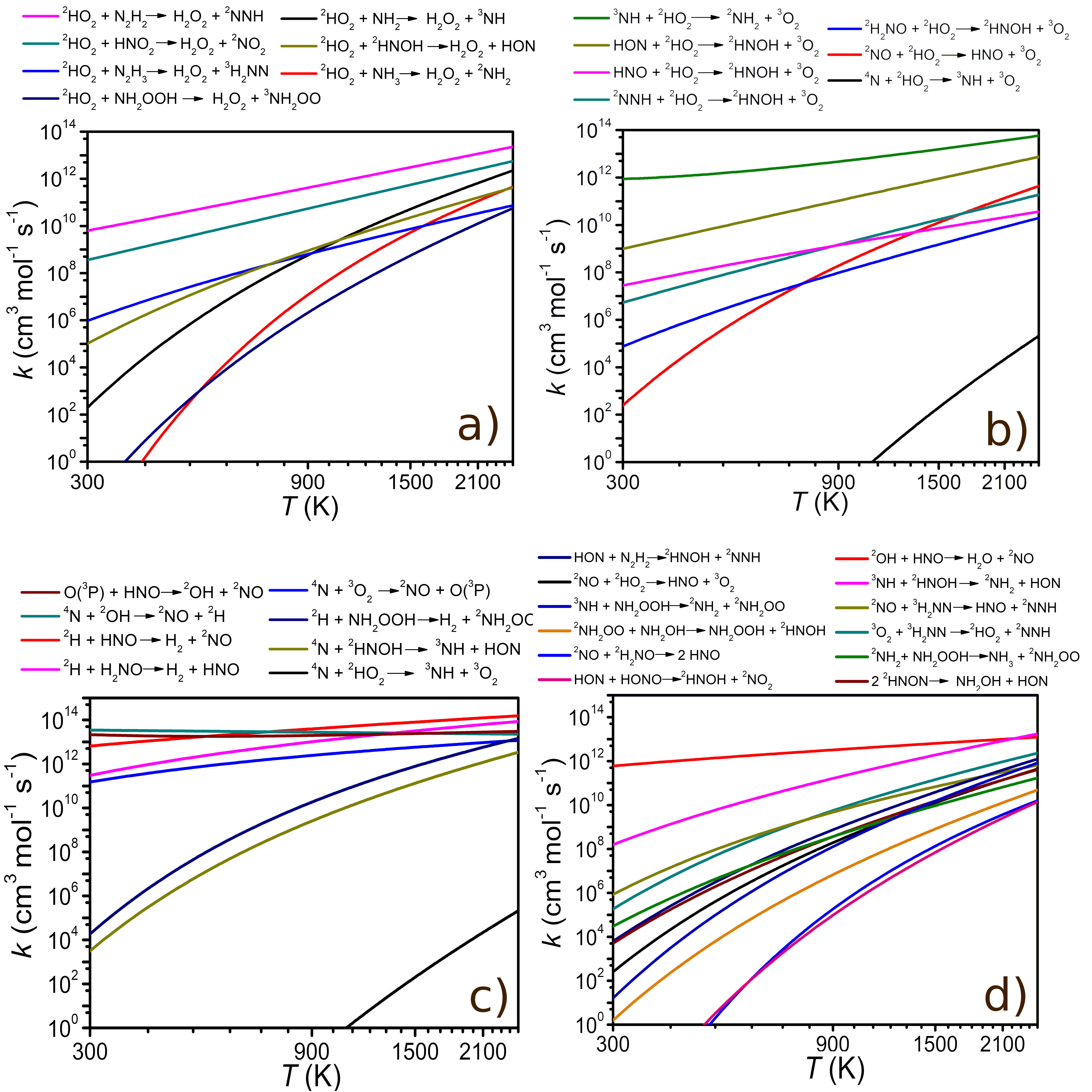}\\
    \caption{The temperature dependencies of the abstraction rate coefficients updated and/or computed in this work.}
        \label{abstractionFig}
\end{figure*}

Master-equation simulations for the
\begin{align}
\ce{^2H + ^2H2NO -> (NH2OH/NH3O) &->  H2 + HNO}  \tag{R8a} \\
\ce{&-> ^2NH2 + ^2OH} \tag{R8b} \\
&\text{and} \notag \\
\ce{^2HO2 + ^3NH -> (^2HNOOH/^2H2NOO) &-> ^2OH + HNO} \tag{R9a} \\
\ce{ &-> ^3O2 + ^2NH2} \tag{R9b} \\
\ce{ &-> ^3O + ^2H2NO} \tag{R9c}
\end{align}
reactions revealed that the initially formed adducts never stabilize, and that these reactions are at the low-pressure limit under all practical conditions ($p \leq 100~\m{bar}$). For R8, we found that R8b is the dominant channel under almost all conditions, but the direct-abstraction reaction \ce{^2H + ^2H2NO -> H2 + HNO} becomes competitive at higher temperatures. The situation is quite similar for R9. Of the well-skipping reactions, only R9a is kinetically significant, but the direct-abstraction reaction to form \ce{^2NH + ^3O2} begins to compete with it at elevated temperatures. The results are depicted graphically in Fig. \ref{H2NOFig} and  Fig. \ref{NHOOHFig}.
For the bimolecular abstraction reactions tabulated in Table \ref{ratecoefficienttable}, the temperature dependencies of the rate coefficients are depicted in Fig. \ref{abstractionFig}.

\subsection{Thermochemistry Calculations}
\begin{figure}[H]
    \centering
    \includegraphics[width=\linewidth]{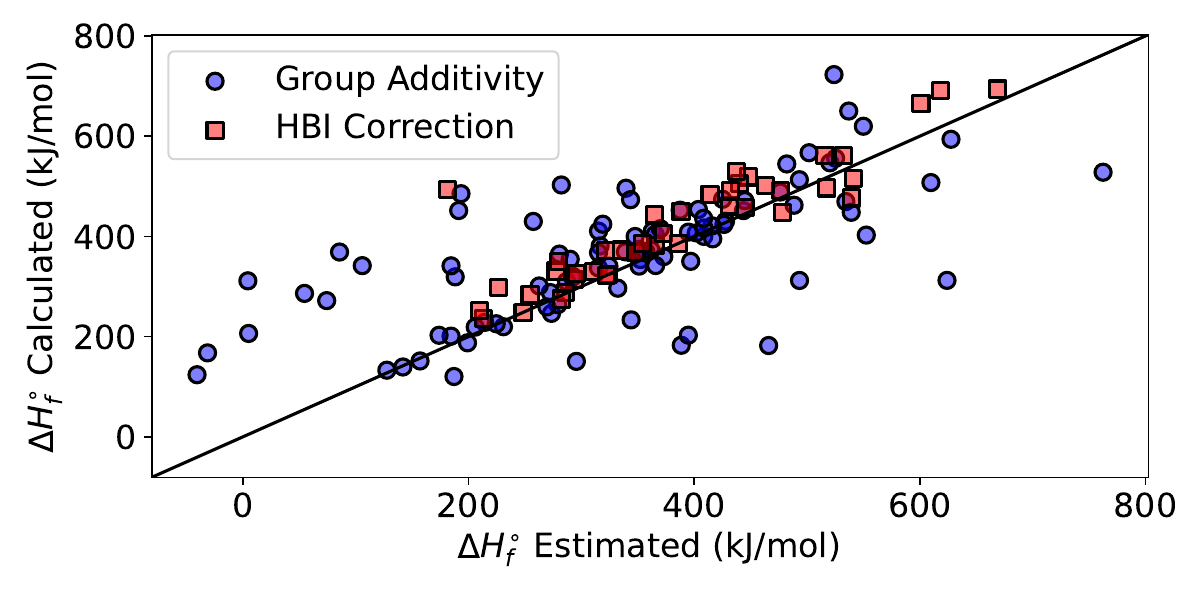}
    \caption{Standard enthalpies of formation ($\Delta_\m{f}H_\m{298~K}^{\ominus}$) calculated at DLPNO-CCSD(T)-F12/cc-pVTZ-F12//$\omega$B97X-D/def2-TZVP are compared to estimates by RMG \cite{RMG2021,RMG2022}, including group additivity (blue circles) and hydrogen bond increment (HBI) corrections \cite{Pang2024} (red squares).}
    \label{fig:edge_thermo_parity}
\end{figure}

Fig. \ref{fig:edge_thermo_parity} shows a parity plot of the standard enthalpy of formation ($\Delta_\m{f}H_{\text{298~K}}^{\ominus}$) estimated by RMG via group-additivity or hydrogen-bond-increment (HBI) correction to quantum chemistry calculations according to Pang et al. \cite{Pang2024} for 137 edge species that passed screening. While most estimates agree within 100~\si{kJ.mol^{-1}}, deviations of several hundred~\si{kJ.mol^{-1}} are observed for some species, particularly those obtained from group-additivity methods. The updated thermochemical data were added to the RMG-database and subsequently used for mechanism generation in this work.

\subsection{Summary of RMG Mechanism} 
\begin{figure}[H]
    \centering
    \includegraphics[width=\linewidth]{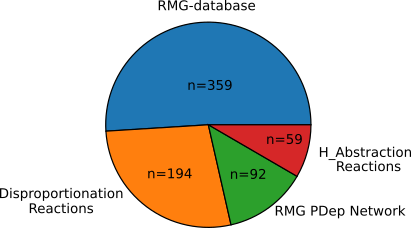}
    \caption{Sources of rate-coefficient parameters in the current RMG core mechanism.}
    \label{fig:rmg_kinetic}
\end{figure}

The RMG generated a mechanism that consists of 53 species and 704 reactions. The corresponding mechanism files in the Cantera and Chemkin formats are provided in the ESI\dag. All thermochemical parameters are taken from high-level calculations or recommendations by Dana et al. \cite{DANA2024} as shown in Fig. \ref{fig:rmg_kinetic}, the kinetic parameters for 359 reactions are sourced from the literature (RMG-database), and 92 are estimated via the pressure-dependent network. The remaining rates are determined using estimation rules, primarily for disproportionation ($n=194$) and hydrogen abstraction ($n=59$) reactions.

\subsection{Model Validation}
Fundamental combustion experiments of ammonia and ammonia/hydrogen mixtures provide valuable insights into how global reactivity of the fuel mixtures and the generation and destruction of chemical species of interest depend on temperature ($T$), pressure ($p$), equivalence ratio ($\phi$), blending ratio, etc. The experimental data also serve as validation targets that are necessary to develop an accurate and predictive combustion mechanism.
The performance of the developed mechanism is validated against a comprehensive set of experimental data for the combustion of pure \ce{H2}, pure \ce{NH3}, and \ce{NH3}/\ce{H2} mixtures with varying \ce{H2} content. This section presents a selection of these validation results, focusing on key combustion targets: ignition delay times (IDTs), laminar-burning velocities (LBVs), and species profiles measured in jet-stirred reactors (JSRs) and flow reactors (FRs). To benchmark its performance, our mechanism is presented in two versions: one that implements the mixtures developed by the Burke group (RMG\_Burke) and one that does not (RMG). Furthermore, we compare the performance of these two models with five recently developed state-of-the-art mechanisms. \cite{ZHU2024113239, STAGNI2023144577, ZHANG2021111653, ZHANG2023127676, MONGEPALACIOS2024101177}

\begin{table*} 
    \centering
    \scriptsize
    \renewcommand{\arraystretch}{1.5}
    \begin{tabular}{|c|c|c|c|c|c|}
    \hline
No. & Reaction & $A$ (\si{cm^3.mol^{-1}.s^{-1}})  & $n$ &  $E_\m{a}$ (\si{kJ.mol^{-1}}) & References \\
         \hline
         \multicolumn{6}{|c|}{} \\ 
    \multicolumn{6}{|c|}{Bimolecular abstraction reactions} \\ \hline 
    
      1 & \ce{^4N + ^2HO2 -> ^3NH + ^3O2} & $1.82\times 10^{-27}$ & 9.81 & 60.5 & \cite{Grinberg2025}\\
      \hline

      2 & \ce{^2NO + ^2HO2 -> HNO + ^3O2} & $1.19\times 10^{-5}$ & 5.06 & 29.9 & \cite{Grinberg2025}\\
      \hline

      3 & \ce{^2H2NO + ^2HO2 -> NH3O + ^3O2} & $1.40\times 10^{-5}$ & 4.53 & 8.1 & \cite{Grinberg2025}\\
      \hline

      4 & \ce{^2NNH + ^2HO2 -> N2H2 + ^3O2} & $6.25\times 10^{-5}$ & 4.57 & 2.2 & \cite{Grinberg2025}\\
      \hline

      5 & \ce{^2HO2 + ^2NH2 -> H2O2 + ^3NH} & $1.97\times 10^{-5}$ & 5.23 & 34.2 & \cite{Grinberg2025}\\
      \hline

      6 & \ce{^2HO2 + NH3 -> H2O2 + ^2NH2} & $4.44\times 10^{-1}$ & 4.00 & 75.5 & \cite{Grinberg2025}\\
      \hline

      7 & \ce{^2HO2 + ^2N2H3 -> H2O2 + ^3H2NN} & $2.79\times 10^{-3}$ & 4.00 & 7.9 & \cite{Grinberg2025}\\
      \hline

      8 & \ce{^2HO2 + HNO2 -> H2O2 +  ^2NO2} & $2.49\times 10^{-3}$ & 4.52 & 0.2 & \cite{Grinberg2025}\\
      \hline

      9 & \ce{^2HO2 + N2H2 -> H2O2 + ^2NNH} & $7.79\times 10^{-1}$ & 3.96 & -0.6 & \cite{Grinberg2025}\\
      \hline

      10 & \ce{HNO + ^2HO2 -> ^2HNOH + ^3O2} & $3.08\times 10^{0}$ & 2.98 & 2.4 & \cite{Grinberg2025}\\
      \hline

      11 & \ce{^2H + HNO -> H2 + ^{2}NO}  & $1.66\times10^{10}$ & 1.18 & 1.87 &  \cite{KLIPPENSTEIN2023} \\
         \hline

      12 & \ce{^{2}OH + HNO -> H2O + ^{2}NO} &$1.20\times10^{9}$ & 1.19 & 1.40 &  \cite{NGUYEN200494} \\
         \hline     

      13 & \ce{^4N + ^3O2 -> ^2NO + O(^3P)} & $5.90 \times 10^{9}$ & 1.01 & 6.28 & \cite{BAULCH2005}\\
      \hline
      
      14 & \ce{^4N + ^2OH -> ^2NO + ^2H} & $1.08 \times 10^{14}$ & -0.20 & 0.0& \cite{BAULCH2005}\\
      \hline

      15 & \ce{^2H + ^{2}H2NO -> H2 + HNO}  & $9.60\times10^8$ & 1.50 & 6.97  & pw (RMG estimate)\\
         \hline
         
      16 & \ce{^{2}NO + ^{2}H2NO -> 2 HNO}  & $1.77 \times 10^{-2}$ & 4.04 & 84.66 & pw \\
         \hline     

      17 & \ce{^3O2 + ^3H2NN -> ^2HO2 + ^2NNH} & $1.16\times 10^{-1}$ & 4.05 & 22.0 &  pw \\
      \hline

      18 & \ce{HON + ^2HO2 -> ^2HNOH + ^3O2} & $7.18\times 10^{-2}$ & 4.13 & 0.5 &  pw \\
      \hline

      19 & \ce{^2HO2 + ^2HNOH -> H2O2 + HON} & $2.56\times 10^{-5}$ & 4.86 & 14.0 &  pw \\
      \hline

      20 & \ce{^4N + ^2HNOH -> ^3NH + HON} & $7.77\times 10^{-2}$ & 4.22 & 33.5 &  pw \\
      \hline

      21 & \ce{^3NH + ^2HNOH -> ^2NH2 + HON} & $1.18\times 10^{1}$ & 3.65 & 11.0 &  pw \\
      \hline

      22 & \ce{^2NO + ^3H2NN -> HNO + ^2NNH} & $1.49\times 10^{1}$ & 3.25 & 18.9 &  pw \\
      \hline

      23 & \ce{HON + N2H2 -> ^2HNOH + ^2NNH} & $7.12\times 10^{-7}$ & 5.50 & 21.1 &  pw \\
      \hline

      24 & \ce{2 ^2HNOH -> NH2OH + HON} & $3.55\times 10^{-6}$ & 5.15 & 20.7 &  pw \\
      \hline

      25 & \ce{HON + HONO -> ^2HNOH + ^2NO2} & $8.07\times 10^{-12}$ & 6.62 & 59.9 &  pw \\
      \hline

      26 & \ce{^2H + NH2OOH -> H2 + ^2NH2OO} & $3.75\times 10^{2}$ & 3.35 & 37.9 &  pw \\
      \hline

      27 & \ce{^2HO2 + NH2OOH -> H2O2 + ^2NH2OO} & $2.37\times 10^{-10}$ & 6.27 & 44.4 &  pw \\
      \hline

      28 & \ce{^2NH2 + NH2OOH -> NH3 + ^2NH2OO} & $4.29\times 10^{-5}$ & 4.69 & 15.9 &  pw \\
      \hline

      29 & \ce{^3NH + NH2OOH -> ^2NH2 + ^2NH2OO} & $9.21\times 10^{-7}$ & 5.51 & 36.8 &  pw \\
      \hline

      30 & \ce{^2NH2OO + NH2OH ->  NH2OOH + ^2HNOH} & $1.31\times 10^{-9}$ & 5.96 & 32.6 &  pw \\
      \hline
      
      31 & \ce{O(^{3}P) + HNO -> ^{2}OH + ^{2}NO} &         $7.24\times10^{15}$ & -1.04 & 0.0644 & pw\\
      &&$2.84\times10^{12}$ & 0.322 & 4.68 & \\ 
       \hline 

      32 & \ce{^3NH + ^2HO2 -> ^2NH2 + ^3O2} & $2.40\times10^{-26}$ & 11.1 & 75.0 & pw\\
      &&$4.34\times10^{3}$ & 2.95 & -5.79 & \\ 
       \hline 

\multicolumn{6}{|c|}{} \\ 
\multicolumn{6}{|c|}{Well-skipping bimolecular reactions (low-pressure limit)} \\ \hline 
    33 & \ce{^{2}NH2 + O(^{3}P) -> (NH2O/NHOH) -> ^{3}NH + ^{2}OH} & $3.09 \times 10^3$ & 2.84 & -11.63 & \cite{KLIPPENSTEIN2023}\\
      \hline

    34 & \ce{^{2}NH2 + O(^{3}P) -> (NH2O/NHOH) -> ^{2}NO + H2}  &$2.38 \times 10^{12}$ & 0.112 & -1.452 & \cite{KLIPPENSTEIN2023}\\
         \hline

    35 & \ce{^{2}NH2 + O(^{3}P) -> (NH2O/NHOH) -> HNO + H} &  $2.78 \times 10^{13}$ & -0.065 & -0.787 & \cite{KLIPPENSTEIN2023}\\
         \hline         

    36 & \ce{H + ^{2}H2NO -> (NH2OH/NH3O) -> H2 + HNO}  &  $8.69\times10^{3}$ & 2.60 & 4.71  & pw \\
         \hline
    37 & \ce{H + ^{2}H2NO -> (NH2OH/NH3O) -> ^2NH2 + ^2OH} & $3.704\times 10^{13}$ & 0.15 & 8.13 & pw \\
         \hline         

    38 & \ce{^2HO2 + ^3NH -> (^2HNOOH) -> ^2OH + HNO} & $2.60\times 10^{7}$ & 1.81 & -10.5 & pw \\
         \hline        

    39 & \ce{^2HO2 + ^3NH -> (^2HNOOH/^2H2NOO) -> ^3O2 + ^2NH2} & $2.22\times 10^{27}$ & -5.36 & 25.6 & pw \\
         \hline     

    40 & \ce{^2HO2 + ^3NH -> (^2HNOOH/^2H2NOO) -> ^3O + ^2H2NO} & $1.24\times 10^{8}$ & 0.386 & 110.7 & pw \\
         \hline        

    \end{tabular}
    \caption {Modified Arrhenius parameters for the reactions added or updated in our mechanism.}
    \label{ratecoefficienttable}
\end{table*}

\subsubsection{Ignition Delay Time}

Ignition delay times (IDTs) measured in shock tubes are essential for validating the intermediate- and high-temperature oxidation mechanisms of ammonia-based fuels. The performance of the current mechanism was evaluated by comparing predicted IDTs against two comprehensive experimental datasets for pure \ce{NH3} and \ce{NH3}/\ce{H2} mixtures. Mathieu and Petersen \cite{mathieu_experimental_2015} measured IDTs of \ce{NH3}/\ce{O2}/\ce{Ar} mixtures over a wide range of conditions, covering 1560--2455 K, pressures of approximately 1.4, 11, and 30 atm, and equivalence ratios of 0.5, 1.0, and 2.0. More recently, Chen et al.\cite{chen_effect_2021-1} investigated the influence of \ce{H2} addition on the autoignition of \ce{NH3} in stoichiometric \ce{NH3}/\ce{H2}/\ce{O2}/\ce{Ar} mixtures. Their measurements span 1020--1945 K, pressures of 1.2 and 10 atm, and \ce{H2} mole fractions from 0\% to 70\%. These datasets provide crucial targets for evaluating the high-temperature performance of our mechanism.

\begin{figure}
    \centering
    \includegraphics[width=0.95\linewidth]{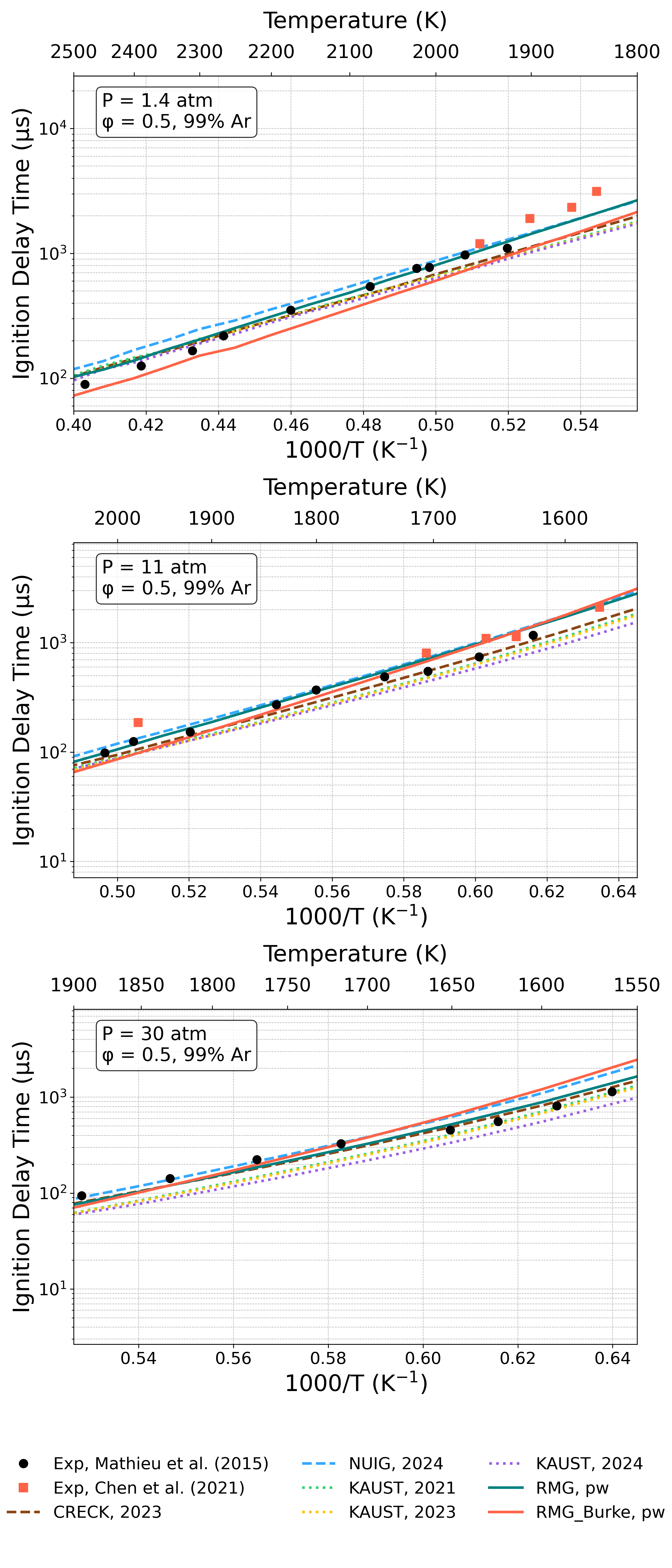}
    \caption{The \ce{NH3} IDT predictions of the current RMG mechanisms and recently developed literature mechanism\cite{ZHU2024113239,STAGNI2023144577,ZHANG2021111653,ZHANG2023127676,MONGEPALACIOS2024101177} compared with experimental data.\cite{mathieu_experimental_2015, chen_effect_2021-1}}
    \label{fig:IDT_NH3}
\end{figure}

Fig. \ref{fig:IDT_NH3}  presents a comparison between model-predicted IDTs and shock tube data for pure \ce{NH3}. At 1.4 atm, most mechanisms underpredict the IDTs measured by Chen et al.,~\cite{chen_effect_2021-1} with the RMG mechanism providing the best agreement, yielding an average error of approximately 12\%. The RMG\_Burke mechanism consistently predicts the shortest ignition delays, whereas the NUIG\_2024 mechanism overpredicts the experimental IDTs across the entire temperature range. At 11~atm, good agreement is observed between most mechanisms and the experimental data from Mathieu et al.~\cite{mathieu_experimental_2015} At 30~atm, most mechanisms underpredict the IDTs at lower temperatures, while at higher temperatures, the experimental values lie approximately midway between the predictions. Overall, the RMG mechanism demonstrates the most robust performance across all conditions, while KAUST\_2024 tends to underpredict and NUIG\_2024 markedly overpredicts ignition delays.

\begin{figure}[H]
    \centering
    \includegraphics[width=1\linewidth]{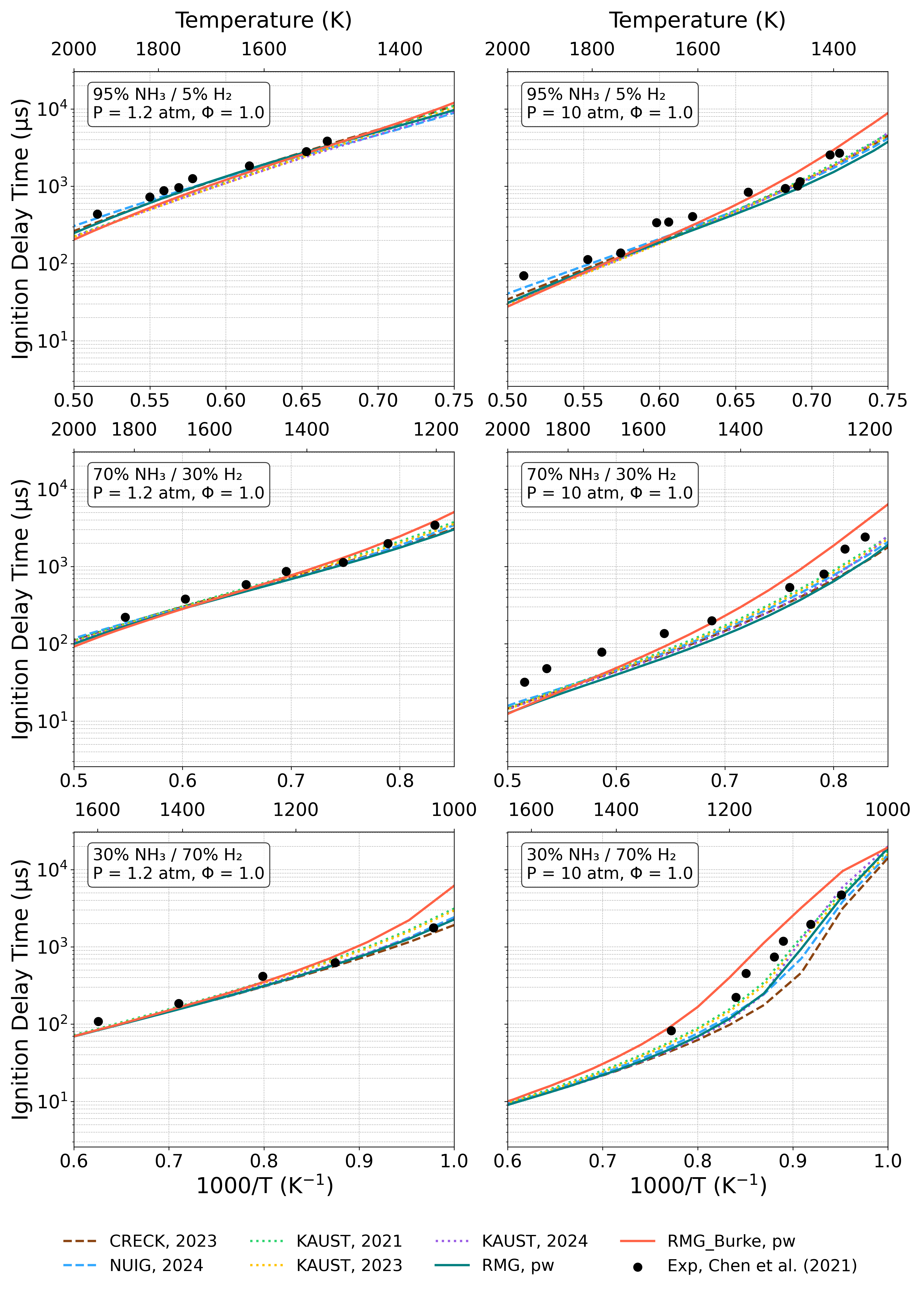}
    \caption{The \ce{NH3}/\ce{H2}  IDT predictions of the current RMG mechanisms and recently developed literature mechanism \cite{ZHU2024113239,STAGNI2023144577,ZHANG2021111653,ZHANG2023127676,MONGEPALACIOS2024101177} compared with experimental data.\cite{chen_effect_2021-1}}
    \label{fig:IDT_NH3_H2}
\end{figure}

The model's performance for the ignition of \ce{NH3}/\ce{H2} mixtures was evaluated against the experimental data of Chen et al.,\cite{chen_effect_2021-1} as shown in Fig.~\ref{fig:IDT_NH3_H2}. At low pressure (1.2~atm), nearly all mechanisms demonstrate excellent agreement with the experimental IDTs. Among them, the RMG and NUIG\_2024 mechanisms perform slightly better, while the RMG\_Burke mechanism is an outlier, tending to overpredict the IDT at lower temperatures ($\sim$1000~K). At higher pressure (10~atm), the deviation between mechanisms increases, and nearly all mechanisms underpredict the experimental IDTs.  The RMG\_Burke is a notable exception, slightly overpredicting reactivity at lower temperatures for the 5\% and 30\% \ce{H2} mixtures, and consistently overpredicting across the full temperature range for the 70\% \ce{H2} blend. In contrast, the KAUST 2021 and NUIG\_2024 mechanisms demonstrate comparatively better agreement under high-pressure conditions. This shared deficiency in underpredicting reactivity suggests that the complex kinetic interactions between \ce{NH3} and \ce{H2} species under \ce{H2}-rich, high-pressure combustion are not yet fully captured by current mechanisms. These discrepancies highlight a critical area for further refinement of high-pressure, \ce{NH3}/\ce{H2} combustion mechanisms.

\subsubsection{Laminar-Burning Velocity}
Laminar-burning velocity (LBV) is a key target to test a model's rate-coefficient parameterisations at high temperatures, particularly for reactions involving hydrogen atoms. It is a fundamental property of a fuel–oxidizer mixture, describing the speed at which a smooth, unstretched flame front propagates through a stationary unburned gas. LBVs are typically measured using a variety of experimental techniques, and multiple datasets under similar conditions allow for assessment of the reliability of these measurements.
For alternative fuels like ammonia and ammonia–hydrogen mixtures, accurate LBV data are vital for designing and optimizing combustion systems for power generation. Recent experimental work by Lhuillier et al.\cite{lhuillier_experimental_2020} provides LBVs of \ce{NH3}/\ce{H2} mixtures at atmospheric pressure, covering variations in unburned gas temperature, hydrogen fraction, and equivalence ratio ($\phi$). These data serve as a comprehensive benchmark for evaluating kinetic mechanisms and exploring the chemical interactions that govern flame propagation.

\begin{figure*}[!t]
    \centering
    \includegraphics[width=1\linewidth]{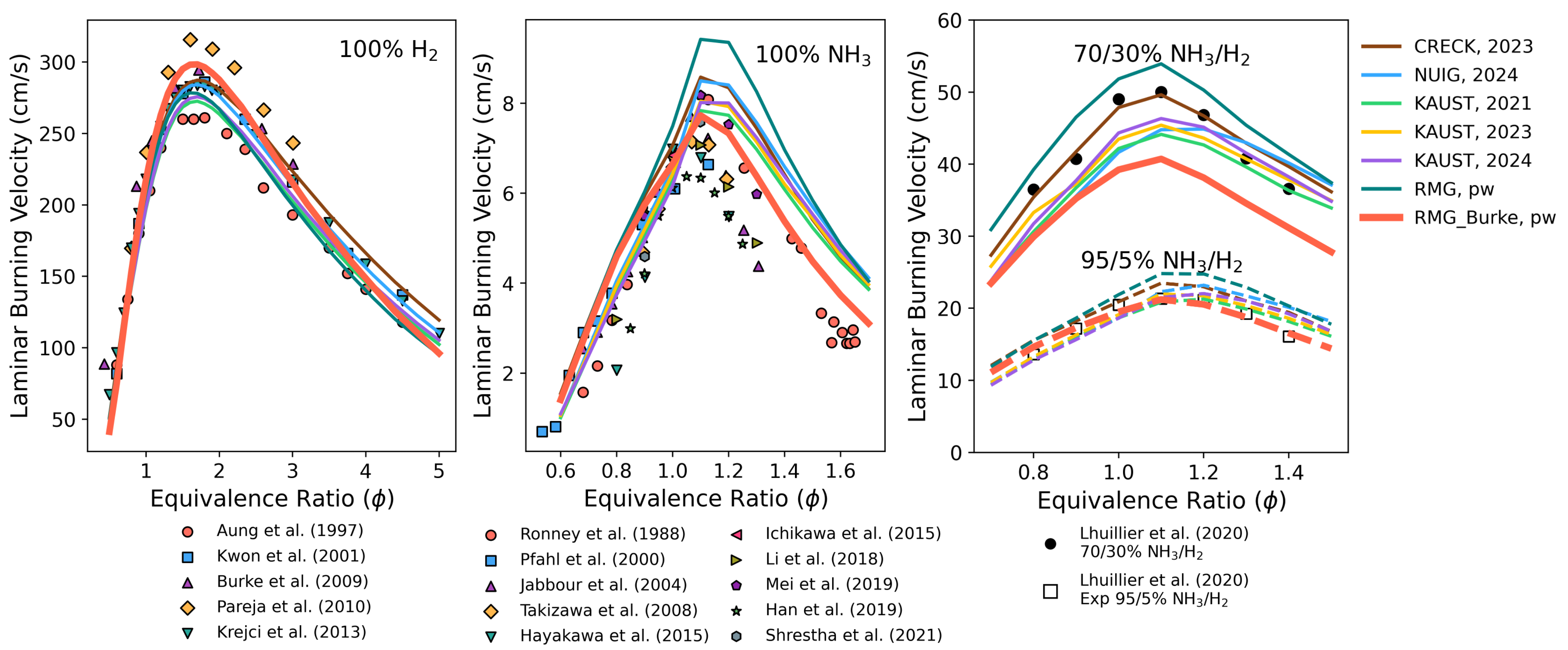}
    \caption{The \ce{NH3}/\ce{H2} LBV predictions of the current RMG mechanisms and recently developed literature mechanism\cite{ZHU2024113239,STAGNI2023144577,ZHANG2021111653,ZHANG2023127676,MONGEPALACIOS2024101177} compared with experimental data for pure \ce{H2},\cite{aung_flame_1997,kwon_flamestretch_2001, burke_effect_2009, pareja_measurements_2010, krejci_laminar_2013} pure \ce{NH3}, \cite{Ronney01051988, PFAHL2000140,jabbour2004,HAYAKAWA201598,TAKIZAWA2008144, ICHIKAWA20159570, Li2018, MEI2019236,HAN2019214, Shrestha2021} and \ce{NH3}/\ce{H2} mixtures. \cite{lhuillier_experimental_2020}}
    \label{fig:LBVSep25_}
\end{figure*}

Fig.~\ref{fig:LBVSep25_} compares experimental and model-predicted LBVs for pure \ce{H2}, pure \ce{NH3}, and \ce{NH3}/\ce{H2} blends with air at atmospheric pressure under different unburned gas temperatures. 
For pure \ce{H2} flames, the results show excellent agreement between experiments and simulations at lean and near-stoichiometric conditions ($\phi$ < 1.5). Hydrogen flames exhibit very high reactivity, with a peak LBV approaching 300 cm/s in the fuel-rich region ($\phi \approx 1.8$). At richer conditions ($\phi$ = 1.5–3), some discrepancies between models emerge. Pareja et al.\cite{pareja_measurements_2010}, using the particle tracking velocimetry (PTV) method, report higher flame speeds, whereas the prediction by Aung\cite{aung_flame_1997}, employing the spherical bomb technique, obtains slightly lower values.  These differences are largely attributable to methodological differences: PTV captures local burning velocities in non-stretched flame regions but may be affected by flame curvature, while the spherical bomb approach measures unsteady flame propagation over short durations, potentially underestimating the LBV. Despite these methodological differences, all mechanisms and experimental datasets show consistent trends, and the theoretical predictions reproduce the high reactivity and peak location of pure hydrogen flames with good accuracy.

In contrast, the LBV of pure \ce{NH3} flames is substantially lower than that of hydrogen, with values typically under 10 cm/s, reflecting the difficulty of combusting ammonia. The peak flame velocity occurs consistently at a slightly fuel-rich equivalence ratio of $\phi \approx 1.1$ across both experimental data and kinetic mechanisms. Most mechanisms accurately capture the flame speed under lean and near-stoichiometric conditions ($\phi \le 1.0$), but they systematically overestimate LBVs in the fuel-rich regime ($\phi$ > 1.0). Among the kinetic mechanisms evaluated, the RMG\_Burke mechanism demonstrates the best overall agreement with experimental data across the entire equivalence ratio range, particularly in the fuel-rich region where other mechanisms deviate significantly. These findings highlight the challenges of ammonia combustion modeling, where \ce{NH_x} chemistry exerts a strong influence on flame propagation.

Blending hydrogen with ammonia substantially increases the LBV, improving flame stability and making the mixture more suitable for practical energy applications. For a 70\%\ce{NH3} / 30\%\ce{H2} mixture, the peak flame velocity consistently occurs at a fuel-rich equivalence ratio of $\phi \approx 1.1$. 
Among the mechanisms tested, CRECK\_2023 provides the closest agreement with the experimental data. This excellent performance can be traced to its slightly adjusted \ce{H2/O2} core kinetics, particularly the key chain-branching step \ce{O2 + H <=> O + OH}, for which the pre-exponential factor was increased relative to the original evaluated value. Such targeted adjustments, together with optimized third-body efficiencies in the \ce{H + OH + M <=> H2O + M} reaction, strengthen the \ce{H2/O2/H/OH} radical-branching loop and consequently enhance the predicted flame-propagation rate—enabling accurate reproduction of the non-linear increase in laminar burning velocity with hydrogen fraction reported by Lhuillier et al.\cite{lhuillier_experimental_2020}. In contrast, the RMG mechanisms were constructed strictly from first-principles and evaluated literature data, without any empirical tuning of rate coefficients. The standard RMG mechanism tends to overestimate flame speeds, whereas RMG\_Burke slightly underestimates them. A similar trend is observed in a 95\%\ce{NH3}/5\%\ce{H2} mixture, where the RMG-generated mechanism still slightly overpredicts the LBV and the RMG\_Burke mechanism underpredicts it. Nevertheless, all mechanisms capture the experimental data reasonably well at this low hydrogen fraction. These results indicate that mechanism discrepancies become more pronounced as the hydrogen content increases.

Overall, RMG\_Burke provides reliable predictions across nearly all conditions, except the 70\%\ce{NH3}/30\%\ce{H2} blend, where it underestimates the measurements. Since all mechanisms except RMG fall below the experimental values in this case, the discrepancy may partly reflect experimental uncertainty. A general trend is also observed: the standard RMG mechanism tends to predict higher LBVs than experiments (except for pure \ce{H2}), whereas the implementation of mixture rules in RMG\_Burke shifts predictions to slightly lower values.

Jet-stirred reactor (JSR) experiments provide valuable benchmark data for validating kinetic mechanisms under well-mixed and steady operating conditions. For ammonia systems, speciation data from JSR experiments are of particular importance, since the selectivity toward nitrogen-containing products (\ce{NO}, \ce{N2O}) strongly influences pollutant formation. In this study, our predictions are compared with the species concentration measurements obtained in the jet-stirred reactor (JSR) experiments of Osipova et al.\cite{osipova_ammonia_2022} for pure \ce{NH3}, and Zhang et al.\cite{ZHANG2021111653} for \ce{NH3}/\ce{H2} mixtures under lean and stoichiometric conditions (equivalence ratios of 0.25 and 1.0) at atmospheric pressure over the temperature range 800--1280~K. Fourier-transform infrared (FTIR) spectroscopy provided quantitative data for major species (\ce{NH3}, \ce{H2O}, \ce{NO}, \ce{N2O}), although homonuclear species such as \ce{N2} and \ce{H2} were not reported.

\subsubsection{Oxidation in a Jet-Stirred Reactor (JSR)}

\begin{figure}[H]
    \centering
    \includegraphics[width=1\linewidth]{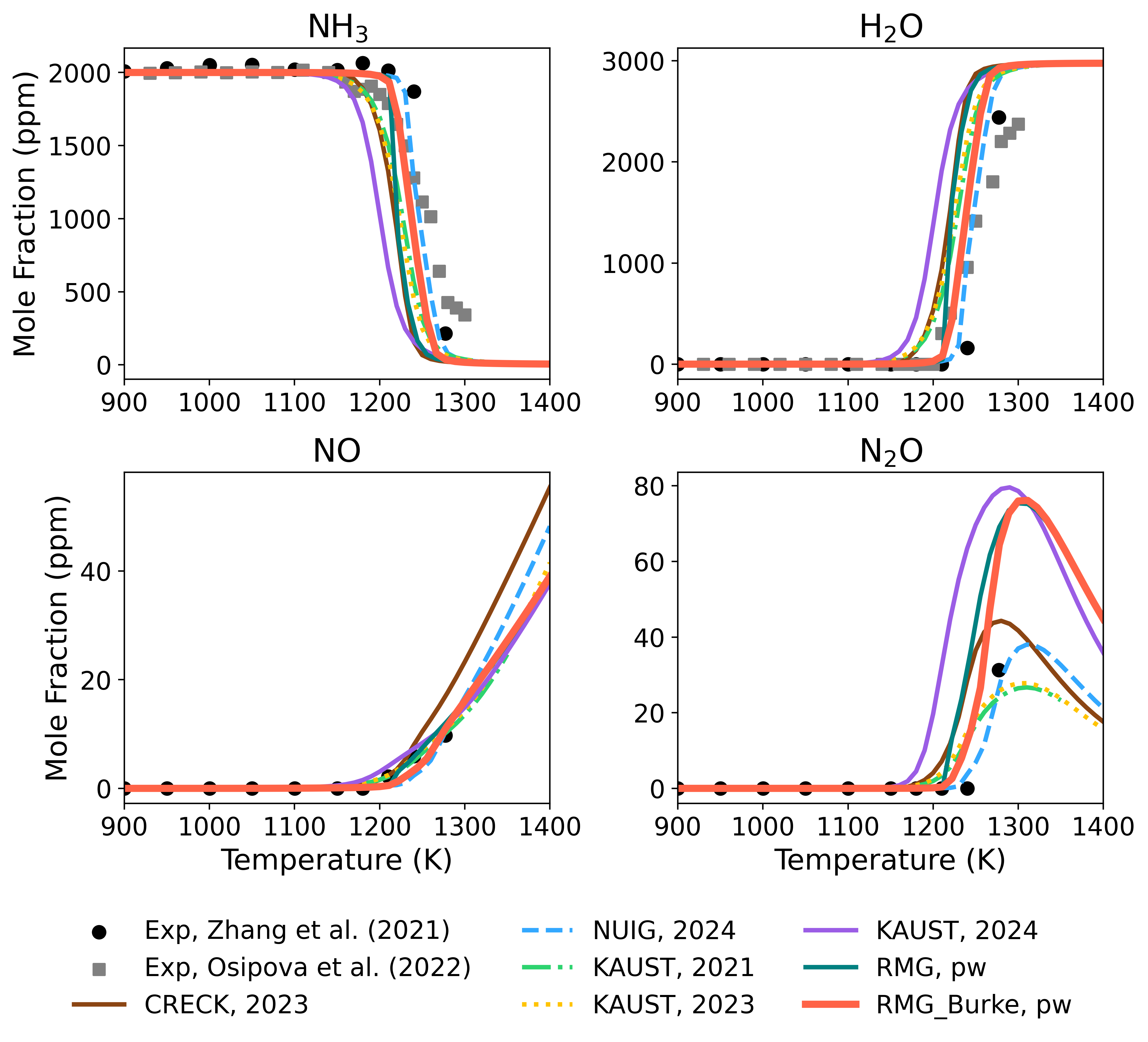}
    \caption{RMG-generated mechanism and representative literature mechanisms \cite{ZHU2024113239,STAGNI2023144577,ZHANG2021111653,ZHANG2023127676,MONGEPALACIOS2024101177} comparisons against species profiles in JSR oxidation of pure \ce{NH3} \cite{ZHANG2021111653, osipova_ammonia_2022}.}
    \label{fig:JSR_NH3}
\end{figure}
\begin{figure}[H]
    \centering
    \includegraphics[width=1\linewidth]{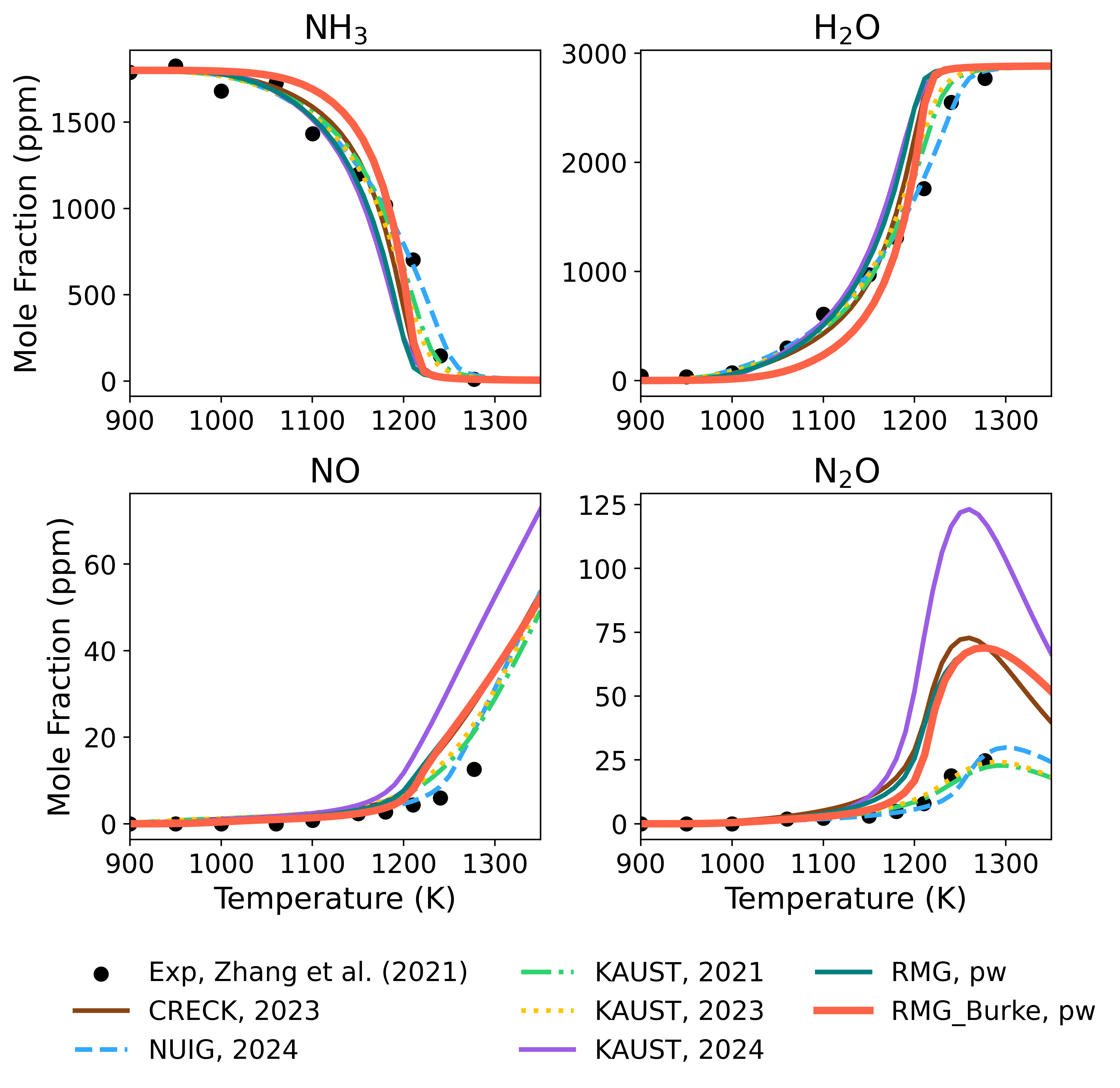}
    \caption{RMG-generated mechanism and representative literature mechanisms \cite{ZHU2024113239,STAGNI2023144577,ZHANG2021111653,ZHANG2023127676,MONGEPALACIOS2024101177} comparisons against species profiles in JSR oxidation of 90\%\ce{NH3}/10\%\ce{H2} mixture \cite{ZHANG2021111653}.}
    \label{fig:JSR_NH3_H2}
\end{figure}

For pure \ce{NH3} oxidation (Fig.\ref{fig:JSR_NH3}), the mechanisms reproduce the primary features of the speciation profiles: \ce{NH3} consumption and the formation of \ce{H2O} and \ce{NO} are captured with good fidelity across the investigated temperature range. The experimental onset temperature for \ce{NH3} consumption is closely matched by RMG, RMG\_Burke, and NUIG\_2024, with \ce{T_{onset}}$\approx 1250~K$. These three mechanisms also show the best agreement for \ce{H2O}, while all mechanisms qualitatively capture the qualitative trend of \ce{NO} formation. The largest model–experiment discrepancies occur for \ce{N2O}, where the experimental scatter is substantial; RMG\_Burke and NUIG\_2024 better reproduce the observed onset than the other mechanisms. However, as the data extend only to $\sim 1280~K$, the relative accuracy of the mechanisms at higher temperatures cannot be firmly assessed.  

For \ce{NH3}/\ce{H2} blends (Fig.~\ref{fig:JSR_NH3_H2}), the onset of reactivity shifts to lower temperatures, with fuel consumption occurring more gradually than in the pure \ce{NH3} case. NUIG\_2024 best reproduces the \ce{NH3} and \ce{H2O} profiles, while RMG\_Burke predicts slightly delayed onsets. For NO, most mechanisms overpredict the onset of formation, and the largest deviations occur for \ce{N2O}, where only KAUST\_2021, KAUST\_2023, and NUIG\_2024 provide reasonable agreement.  For both pure \ce{NH3} and \ce{NH3}/\ce{H2}, KAUST\_2024 consistently predicts earlier transitions in all four major species, systematically overestimating reactivity. Overall, the NUIG\_2024 mechanism achieves the best balance among the tested models but relies on multiple targeted rate adjustments to reproduce the experiment, whereas the RMG-based models retain the original ab initio or measured parameters without fitting to specific datasets.

It is noteworthy that the KAUST\_2024 predicts higher \ce{N2O} yields than KAUST\_2023 under oxygen-lean conditions (1–10~\%), mainly due to the 50\% higher rate coefficient for the \ce{NH + NO -> N2O + H} reaction. In contrast, at higher \ce{O2} concentrations, \ce{N2O} formation proceeds via \ce{NH2 + NO2 -> N2O + H2O}, for which KAUST\_2024 adopts a rate coefficient about half that of KAUST\_2023, resulting in lower predicted \ce{N2O} yields. These differences reflect the competing effects of \ce{NH2}- and \ce{NOx}-related pathways on \ce{N2O} selectivity.

\begin{figure*}[!t]
    \centering
    \includegraphics[width=0.8\linewidth]{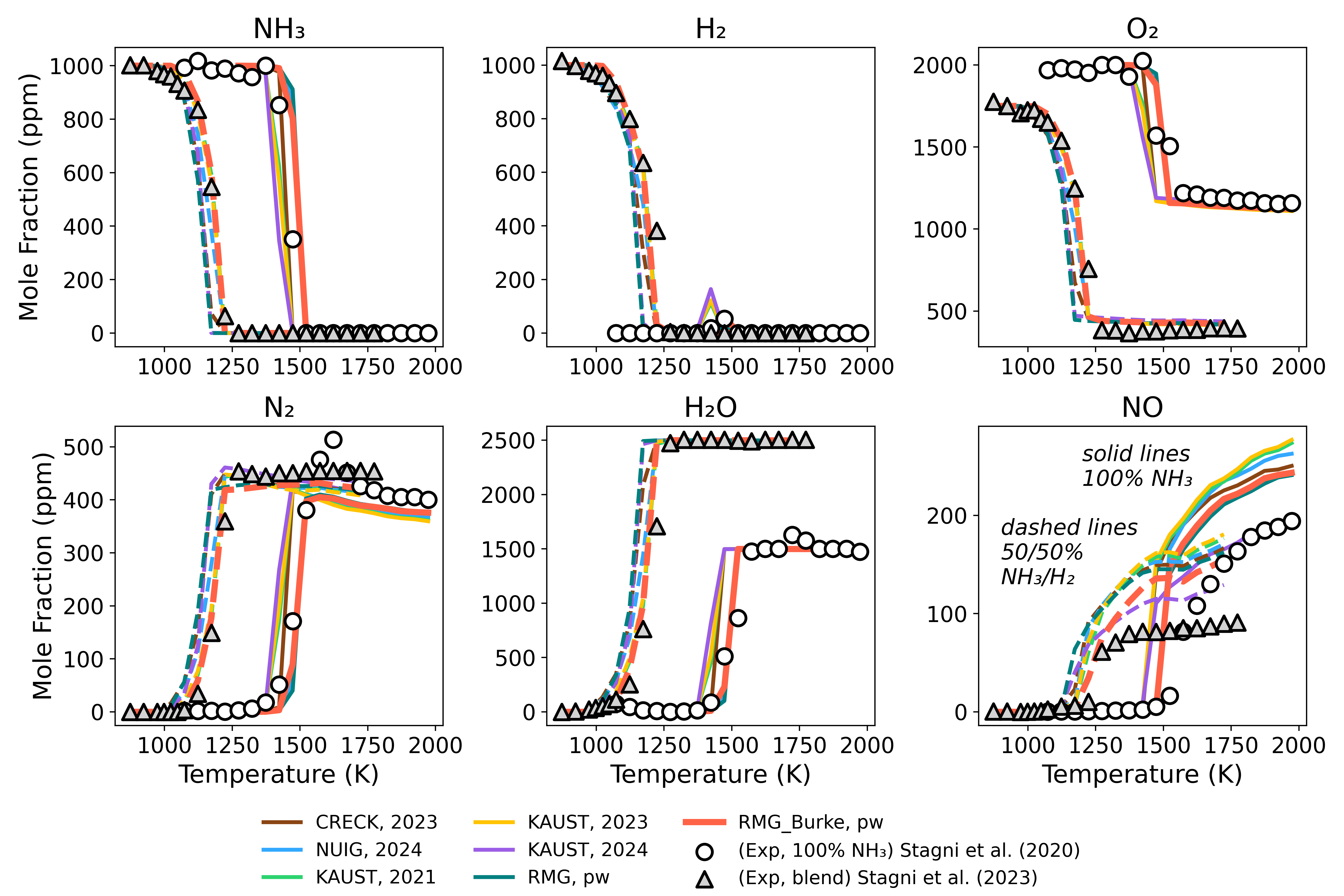}
    \caption{RMG-generated RMG-generated mechanism and representative literature RMG-generated mechanisms \cite{ZHU2024113239, STAGNI2023144577, ZHANG2021111653, ZHANG2023127676, MONGEPALACIOS2024101177} comparisons against species profiles in the FR oxidation; pure \ce{NH3} \cite{STAGNI2020} (solid line) and \ce{50\%NH3}/\ce{50\%H2} mixture \cite{STAGNI2023144577} (dashed line).}
    \label{fig:FR_NH3_H2}
\end{figure*}
\subsubsection{Oxidation in a Flow Reactor (FR)}
Flow-reactor (FR) experiments provide critical data on the intermediate-temperature oxidation behavior of pure \ce{NH3} and \ce{NH3}/\ce{H2} mixtures under short residence times ($\approx50~ms$). In this work, FR simulations were performed using a plug-flow reactor (PFR) model with temperature profiles imposed according to measurements from thermocouples.

Fig. \ref{fig:FR_NH3_H2}, compares simulated and experimental species mole fractions for pure \ce{NH3} and a 1:1 \ce{NH3}/\ce{H2} mixture, alongside predictions by literature mechanisms \cite{STAGNI2020, STAGNI2023144577, ZHU2024113239, ZHANG2021111653, ZHANG2023127676, MONGEPALACIOS2024101177}. Our kinetic mechanism effectively reproduces reactant consumption (\ce{NH3}, \ce{O2}, \ce{H2}) and the formation of major combustion products (\ce{N2}, \ce{H2O}). Across all conditions, the largest discrepancies are observed for \ce{^2NO}, which is generally overestimated. Among the tested mechanisms, NUIG\_2024 again yields the closest agreement for both \ce{NO} and \ce{N2O} under FR conditions, benefiting from its empirically refined rate-coefficient parameters. At the same time, RMG\_Burke shows comparable performance using unmodified literature values. KAUST\_2024 predicts noticeably lower \ce{^2NO} concentrations, reflecting differences in the treatment of key reactions producing or consuming \ce{^2NO} (reactions 13-14, Table 1). These results indicate that \ce{^2NO}-forming pathways require further investigation for accurate predictive modeling.

Overall, most mechanism predictions are in good agreement with experimental data. RMG\_Burke provides the most consistent performance across the range of conditions without parameter adjustments, whereas NUIG\_2024 achieves slightly better quantitative agreement through selective rate tuning, particularly in reproducing the experimental temperature dependence of \ce{NO_x} formation.  In contrast, the standard RMG-generated mechanism tends to predict earlier reactivity for most conditions, particularly at higher hydrogen fractions.

\section*{Conclusions}
In this work, we developed with RMG a new combustion mechanism for \ce{NH3}/\ce{H2} blends. Thermochemical and kinetic parameters were updated for many species and reactions with literature values, and when such data were not available, high-level calculations were used to determine them. For six key reactions, we used the mixture rules developed by the Burke group to improve our treatment of bath-gas-composition effects. Importantly, no parameters were tuned to obtain better agreement with experimental data.

Extensive validation against IDTs, LBVs, and species evolution in FR and JSR shows that the mechanism reproduces a broad range of experimental observations with strong predictive capability. In particular, the 
incorporation of improved treatment of mixture effects
provides a more accurate description of the kinetics of pressure-dependent reactions in combustion environments. This improved treatment was generally found to result in better agreement with the experiment, but not always. In the cases where worse agreement was observed, one might need to reassess the rate-coefficient parameterisations of key reactions. We further note that in many models, rate-coefficient parameters are empirically adjusted to obtain better agreement with experiment. However, in some cases, these adjustments could be introduced to correct for the model's failure to treat mixture effects. If so, it is likely that such adjustments will improve the model's performance only under a limited set of experimental conditions and degrade the performance elsewhere. 

Despite this progress, challenges remain. Discrepancies in ignition delay under \ce{H2}-rich, high-pressure conditions and the overprediction of \ce{NO} formation highlight the need for refinement of the \ce{NOx} sub-mechanism. Overall, this work presents a reliable and untuned mechanism for \ce{H2}/\ce{NH3} combustion that explicitly treats bath-gas mixture effects. In future work, the bath-gas-mixture dependence should be expanded to as many reactions as feasible, after which the sensitivities of the key reactions should be reassessed. Given that the error introduced by failing to treat bath-gas-mixture dependence often equals or exceeds the uncertainties in pure-bath-gas rate coefficients, it seems clear that only trying to reduce the uncertainties of the latter will result in very limited gains.

\section*{Author contributions}
Yu-Chi Kao: literature review, kinetic modeling, mechanism generation, writing and editing manuscript. Anna C. Doner: rate and thermochemistry calculations, writing original draft. Timo T. Pekkanen: master-equation rate calculations and writing manuscript. Chuangchuang Cao: literature review and kinetic modeling. Sunkyu Shin: kinetic modeling and editing manuscript. Alon Grinberg Dana: literature review, rate and thermochemistry calculations.
Yi-Pei Li: funding acquisition and supervision.
William H. Green: funding acquisition, supervision, and editing manuscript.

\section*{Conflicts of interest}
There are no conflicts of interest to declare.


\section*{Acknowledgements}
YCK, ACD, TTP, CCC, SKS, and WHG gratefully acknowledge support from ExxonMobil. YCK is supported by the Graduate Students Study Abroad Program (114-2917-I-002-025) sponsored by the National Science and Technology Council in Taiwan. CCC gratefully acknowledges fellowship support from Shanghai Jiao Tong University. TTP acknowledges the Finnish Foundation for Technology Promotion for fellowship funding and CSC IT Center for Science in Finland for computational resources. The authors acknowledge Professor Markus Kraft for his valuable guidance in the preparation of this manuscript and Akshat Shirish Zalte for his assistance with the early-stage calculations.


\bibliography{rsc} 
\bibliographystyle{rsc} 

\end{document}